\documentclass[%
 preprint,
 superscriptaddress,
 amsmath,amssymb,
 aps,
 showkeys,
 titlepage,
 endfloats*.   
]{revtex4-2}

\usepackage[utf8]{inputenc}
\usepackage[english]{babel}

\usepackage{graphicx}
\usepackage{dcolumn}
\usepackage{bm}

\usepackage[colorlinks = true,
            linkcolor = blue,
            urlcolor  = blue,
            citecolor = red,
            anchorcolor = blue]{hyperref}

\begin{document}


\title{Coupled Spin-lattice Dynamics across a Magnetostructural Phase Transition} 

\author{Lokanath Patra}
\affiliation{Department of Mechanical Engineering, University of California, Santa Barbara, CA 93106, USA}

\author{Zeyu Xiang}
\affiliation{Department of Mechanical Engineering, University of California, Santa Barbara, CA 93106, USA}

\author{Yubi Chen}
\affiliation{Department of Mechanical Engineering, University of California, Santa Barbara, CA 93106, USA}
\affiliation{Department of Physics, University of California, Santa Barbara, CA 93106, USA}

\author{Bolin Liao}
\email{bliao@ucsb.edu} \affiliation{Department of Mechanical Engineering, University of California, Santa Barbara, CA 93106, USA}

\date{\today}

\begin{abstract}
First-order magnetostructural phase transitions underpin giant magnetocaloric effects, yet the microscopic role of lattice dynamics in these transitions remains controversial. Here, we use first-principles spin–lattice dynamics simulations to investigate the coupled evolution of magnetization and phonon dispersions across the magnetostructural transition in MnAs. Our simulations quantitatively reproduce the experimentally observed Curie temperature, lattice contraction, and free-energy crossing between hexagonal and orthorhombic phases. We show that below the Curie temperature, magnetic-field–induced hardening of soft phonon modes gives rise to a sizable lattice entropy contribution that enhances the total isothermal entropy change by approximately 23\% under a 5 T field. In contrast, the lattice entropy change associated with the structural phase transition itself has an opposite sign and partially compensates the lattice contributions due to field-induced phonon hardening. This competition reconciles long-standing discrepancies in the interpretation of magnetocaloric entropy measurements across first-order transitions. Additionally, we demonstrate that the strong magnetic-field dependence of the phonon spectrum near the transition enables large tunability of lattice thermal conductivity, highlighting MnAs as a promising platform for magnetic-field-controlled thermal switching. Our results establish a unified microscopic picture of spin–lattice coupling in first-order magnetocaloric materials and provide design principles for engineering enhanced caloric and thermal transport functionalities. 
\end{abstract}

                            
\maketitle


\section{Introduction}
Magnetocaloric effect (MCE) refers to the reversible entropy change of a magnetic material as a function of an external magnetic field~\cite{pecharsky1999magnetocaloric}. MCE has been extensively studied for efficient and environmentally friendly solid-state cooling applications~\cite{lyubina2017magnetocaloric}. Microscopically, MCE stems from the change of magnetization with the external magnetic field due to aligned local magnetic moments at a constant temperature and the associated isothermal entropy change $\Delta S_T$. An intuitive picture is that the increased alignment of the magnetic moments under the magnetic field leads to a more ordered state with a lower magnetic entropy. For this reason, MCE is maximized near a magnetic transition, where the magnetization is the most sensitive to the external magnetic field~\cite{de2010theoretical}.  

While a majority of the magnetocaloric entropy change is attributed to the magnetic order, there is increasing recognition that a sizable additional contribution can come from the crystal lattice~\cite{lazewski2010phonon,landers2018determining,cedervall2019magnetocaloric,Patra2023PhysicalReviewLetters}. This is because the properties of the dynamical vibrational modes of the lattice (phonons) can be sensitively influenced by the magnetic state through spin-lattice coupling. In other words, the entropy of the phonons can also be tuned by an external magnetic field. For example, we recently showed how the phonon dispersion relation in gadolinium (Gd), a good MCE material with strong spin-lattice coupling, evolves as a function of an external magnetic field, and that the lattice contributes roughly 35\% of the total isothermal entropy change at the Curie temperature $T_C$~\cite{Patra2023PhysicalReviewLetters}. This large contribution was attributed to the long-range spin-lattice coupling mediated by the Ruderman-Kittel-Kasuya-Yosida (RKKY) exchange interaction. Based on this physical picture, we expected that the lattice contribution to the entropy change should be even higher in the so-called ``first-order magnetocaloric materials'', which feature first-order magnetostructural phase transitions (concurrent magnetic and structural transitions). Giant MCE has been observed in representative first-order materials such as Gd$_5$(Si$_2$Ge$_2$)~\cite{pecharsky1997giant} and MnAs~\cite{wada2001giant}. However, the exact role played by phonons for the giant MCE in first-order magnetocaloric materials remains under debate~\cite{hao2020sign}.  

MnAs is an ideal model system for studying the magnetostructural transition because of its relatively simple chemistry and crystal structure. MnAs undergoes a first-order ferromagnetic to paramagnetic transition at a Curie temperature $T_C$ = 318 K, which is accompanied by a structural transition from a hexagonal NiAs-type structure (space group P63/mmc) to an orthorhombic MnP-type structure (space group Pnma) with a 2\% volume contraction~\cite{wilson1964crystal}, as illustrated in Fig.~\ref{fig:transition}a. This first-order magnetostructural transition is a direct consequence of spin-lattice coupling, leading to a strong dependence of the exchange interaction on the lattice spacing, which has been successfully modeled using phenomenological theories~\cite{bean1962magnetic-ec1,pytlik1985magnetic-aee}.  A giant $\Delta S_T$ of -30 J\,kg$^{-1}$\,K$^{-1}$ under a 5-T magnetic field near $T_C$ has been experimentally found~\cite{wada2001giant}. The origin of this large $\Delta S_T$ value has been a subject of debate. Mean-field approximation suggests that the majority of $\Delta S_T$ is contributed from the magnetic degree of freedom~\cite{zou2008giant-ba3}. {\L}a{\.z}ewski \textit{et al.} used density functional theory (DFT) to calculate the phonon dispersion of the hexagonal phase of MnAs as a function of the magnetic moment of Mn ions and external pressure~\cite{lazewski2010phonon}. They showed that the phonon dispersion changes drastically with the Mn magnetic moment, indicating a strong spin-lattice coupling.  Since their calculation was performed at 0 K, they used external pressure to mimic the volume contraction at the magnetostructural transition and showed significant softening of an acoustic phonon mode at the $M$ point, which is identified as a driving force for the phase transition. By comparing the phonon dispersions of the hexagonal phase and the orthorhombic phase at the transition pressure, they concluded that the phonon entropy change is 9.31 J\,kg$^{-1}$\,K$^{-1}$ with an opposite sign of the magnetic entropy change. However, they also acknowledged that their calculation did not consider the magnetic-field dependence of the phonon dispersion, thus only partially capturing the phononic entropy change. Furthermore, the paramagnetic nature of the orthorhombic phase was not accurately incorporated in their calculation.

In this work, we conducted finite-temperature first-principles spin-lattice dynamics simulations to explicitly examine the evolution of magnetization and phonon dispersion of MnAs across the magnetostructural transition at $T_C$, moving beyond the 0-K ground-state simulations by {\L}a{\.z}ewski \textit{et al.}~\cite{lazewski2010phonon}.
Using exchange interaction parameters extracted from first-principles calculations, our simulations explicitly reveal how the magnetic structure and the lattice dynamics change as a function of temperature and magnetic field. From our simulations, we clarify that there are two distinct lattice contributions with opposite signs to the total isothermal entropy change: one from magnetic-field-dependent phonon dispersions and the other from the structural phase transition. Furthermore, we demonstrate that the sensitive dependence of phonon properties on magnetic field near the magnetostructural phase transition can be exploited for applications that require dynamic control of thermal transport. Our results provide a complete picture of the role that the lattice plays in first-order magnetocaloric materials, which can guide rational design and discovery of magnetocaloric materials with enhanced properties.

\section{Results and Discussions}
\begin{figure}
    \centering
    \includegraphics[width=0.9\linewidth]{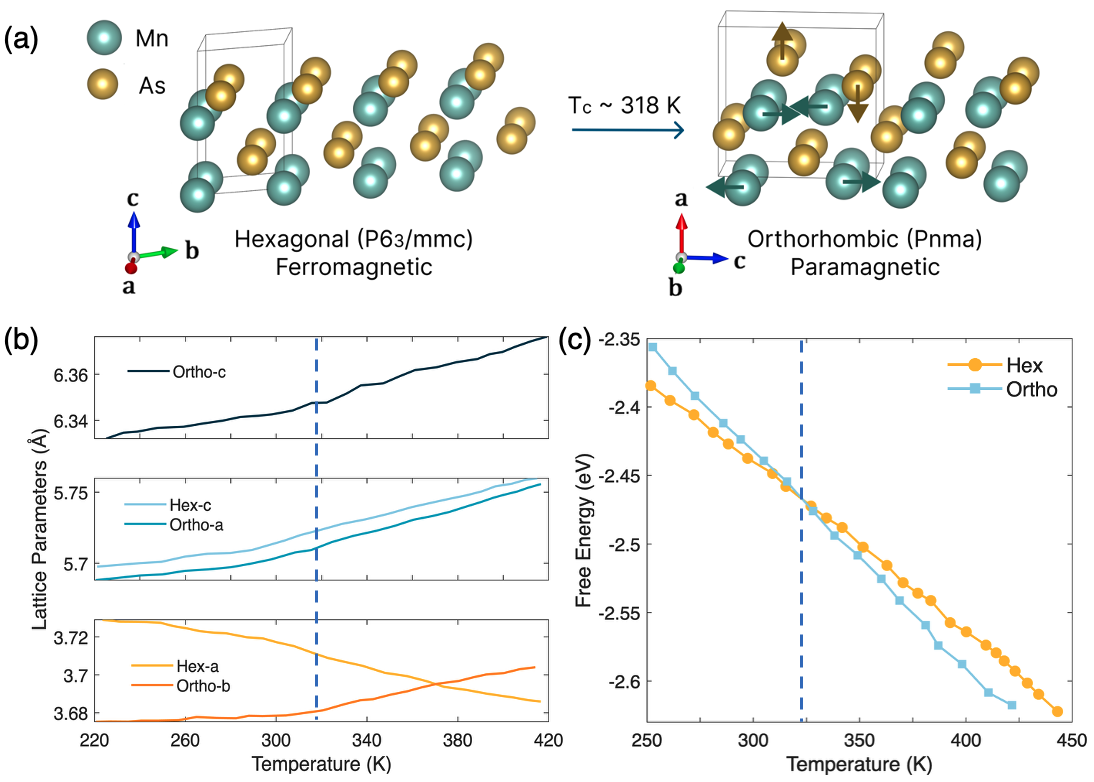}
    \caption{\textbf{Magnetostructural transition in MnAs.} (a) Schematic illustrating the structural transition in MnAs from the hexagonal phase to the orthorhombic phase at $T_C$ of 318 K. Arrows in the orthorhombic phase highlight the atomic distortions during the transition. (b) Simulated lattice constants of both hexagonal (``Hex'') and orthorhombic (``Ortho'') phases of MnAs along different directions as a function of temperature. The directions are labeled according to the convention shown in (a). The blue dashed line marks the experimental $T_C$ of 318 K. (c) Simulated Gibbs free energy of both hexagonal and orthorhombic phases of MnAs as a function of temperature. The blue dashed line marks the crossing that occurs around 323 K, close to the experimental $T_C$ of 318 K.}
    \label{fig:transition}
\end{figure}

Computational details can be found in the Supplementary Information and briefly summarized here. We simulated the temperature-dependent spin and lattice dynamics of the ferromagnetic hexagonal phase of MnAs using SLD. SLD is similar to molecular dynamics (MD) simulations but also explicitly incorporates spin-lattice coupling by considering the dependence of the spin exchange interaction parameters on the interatomic distances at each simulation step~\cite{ma2008large,wu2018magnon,hellsvik2019general,perera2017collective}. Fluctuations of both the local magnetic moments and the atomic positions are simulated in SLD. For the paramagnetic orthorhombic phase of MnAs above $T_C$, we used MD to simulate its lattice dynamics without considering short-range magnetic correlations~\cite{Bocarsly2020Magnetostructural}. Our simulation results are in excellent agreement with experiments, suggesting that the short-range magnetic correlations have negligible impact on the thermodynamic properties of orthorhombic MnAs. The simulated temperature-dependent lattice constants of both phases are shown in Fig.~\ref{fig:transition}b. For the hexagonal phase near $T_C$ (318 K), our simulated lattice parameters are $c$ = 5.72 \AA~ and $a$ = 3.71 \AA. For the orthorhombic phase near $T_C$, the simulated lattice parameters are $a$ = 5.71 \AA, $b$ = 3.68 \AA, and $c$ = 64 \AA. The corresponding crystallographic directions are labeled in Fig.~\ref{fig:transition}a following convention in the literature. These simulated lattice parameters are in excellent agreement with experimental X-ray diffraction results~\cite{willis1954magnetic,wilson1964crystal} and show a significant improvement over ground-state DFT calculations~\cite{lazewski2010phonon,Bocarsly2020Magnetostructural}. In particular, our simulation confirms the significant lattice contraction in the hexagonal plane by 1\% and the volume contraction by 2.3\% during the transition. It is also interesting to note that all three lattice parameters of the paramagnetic orthorhombic phase increase with temperature as a result of thermal expansion, while the volume change of the hexagonal phase is anisotropic: The $c$-axis expands with temperature while the $a$-axis contracts, in agreement with experiments~\cite{willis1954magnetic,suzuki1982relation}. This is a direct consequence of spin-lattice coupling and stems from the competition between the energies associated with interatomic magnetic exchange and chemical bonding~\cite{Bocarsly2020Magnetostructural}. Figure~\ref{fig:transition}c shows the simulated Gibbs free energy of both phases at ambient pressure as a function of temperature. The Gibbs free energy contains contributions from both spin and lattice degrees of freedom in the hexagonal phase and only lattice degree of freedom in the orthorhombic phase. The crossing of the Gibbs free energy of the two phases occurs around 323 K, which is close to the experimental $T_C$ of 318 K, further validating our simulation results and justifying the omission of short-range magnetic correlations in the paramagnetic orthorhombic phase (when evaluating thermodynamic properties).

\begin{figure}
    \centering
    \includegraphics[width=1\linewidth]{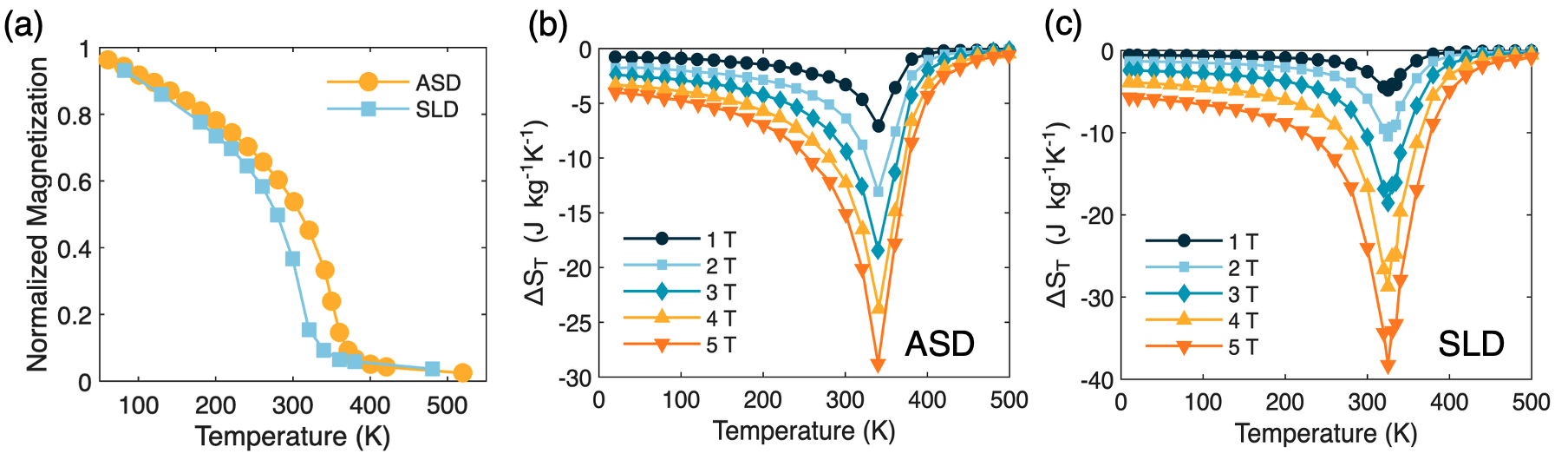}
    \caption{\textbf{Simulated magnetization and isothermal entropy change in hexagonal MnAs.} (a) Simulated zero-field magnetization of hexagonal MnAs as a function of temperature using both ASD and SLD. Simulated isothermal entropy change in hexagonal MnAs as a function of temperature and magnetic field using (b) ASD and (c) SLD. }
    \label{fig:entropy}
\end{figure}

Next, we examine how magnetization in MnAs changes with temperature and the associated isothermal entropy change $\Delta S_T$, which can be evaluated using the thermodynamic Maxwell relation~\cite{pecharsky1999magnetocaloric}:
\begin{equation}
\Delta S (T, \Delta H) = \mu_0 \int_{H_i}^{H_f} \left( \frac{\partial M(T,H)}{\partial T} \right)_H\ dH,
\label{eqn:entropy_change}
\end{equation}
where $\Delta H = H_f - H_i$ is the change of the applied external field ($H_f$ and $H_i$ are final and initial fields, respectively), $\mu_0$ is the Bohr magneton, $M$ is the magnetization and $T$ is the temperature.
To observe separate contributions from spin and lattice degrees of freedom, here we compare the results of atomistic spin dynamics (ASD) and SLD simulations of the hexagonal phase of MnAs~\cite{Patra2023PhysicalReviewLetters}. In ASD, the atomic positions are fixed and only the local magnetic moments are allowed to fluctuate~\cite{antropov1995ab,skubic2008method,ma2014dynamic,patra2024impact,yuan2024enhancing}. In this case, the evolution of the magnetization and the isothermal entropy change $\Delta S_T$ originate entirely from the magnetic degree of freedom. In comparison, SLD simulates the coupled dynamics of local magnetic moments and atomic positions and thus contains the contributions from both magnetic and lattice degrees of freedom. Figure~\ref{fig:entropy}a shows the zero-field magnetization versus temperature in hexagonal MnAs simulated by both ASD and SLD. The result suggests that ASD overestimates the Curie temperature, while the SLD prediction is closer to the experimental value. Moreover, the SLD result shows a sharper transition of the magnetization near the Curie temperature, indicating a higher $\Delta S_T$ as a result of the lattice contribution, according to Eqn.~\ref{eqn:entropy_change}. We note here that our SLD simulation did not capture the discontinuous change of magnetization at a first-order phase transition because the simulation shown here was performed only on the hexagonal phase. $\Delta S_T$ can be calculated from the temperature- and field-dependent magnetization based on Eqn.~\ref{eqn:entropy_change} (see the Supplementary Information), and the results from ASD and SLD are shown in Fig.~\ref{fig:entropy}b and \ref{fig:entropy}c, respectively. With a magnetic field of 5\,T, the maximum $\Delta S_T$ predicted from ASD is roughly -30 J\,kg$^{-1}$\,K$^{-1}$. This value seemingly agrees with the experimental value; however, as pointed out previously, the ASD simulation only includes a contribution from the magnetic degree of freedom ($\Delta S_M$). In comparison, the SLD simulation predicts a maximum $\Delta S_T$ of roughly -39 J\,kg$^{-1}$\,K$^{-1}$ with a magnetic field of 5\,T. This result suggests a contribution from the lattice of approximately -9 J\,kg$^{-1}$\,K$^{-1}$ ($\Delta S_L$), about 23\% of the total entropy change. This value is approximate because ASD and SLD predict different Curie temperatures.

\begin{figure}
    \centering
    \includegraphics[width=0.8\linewidth]{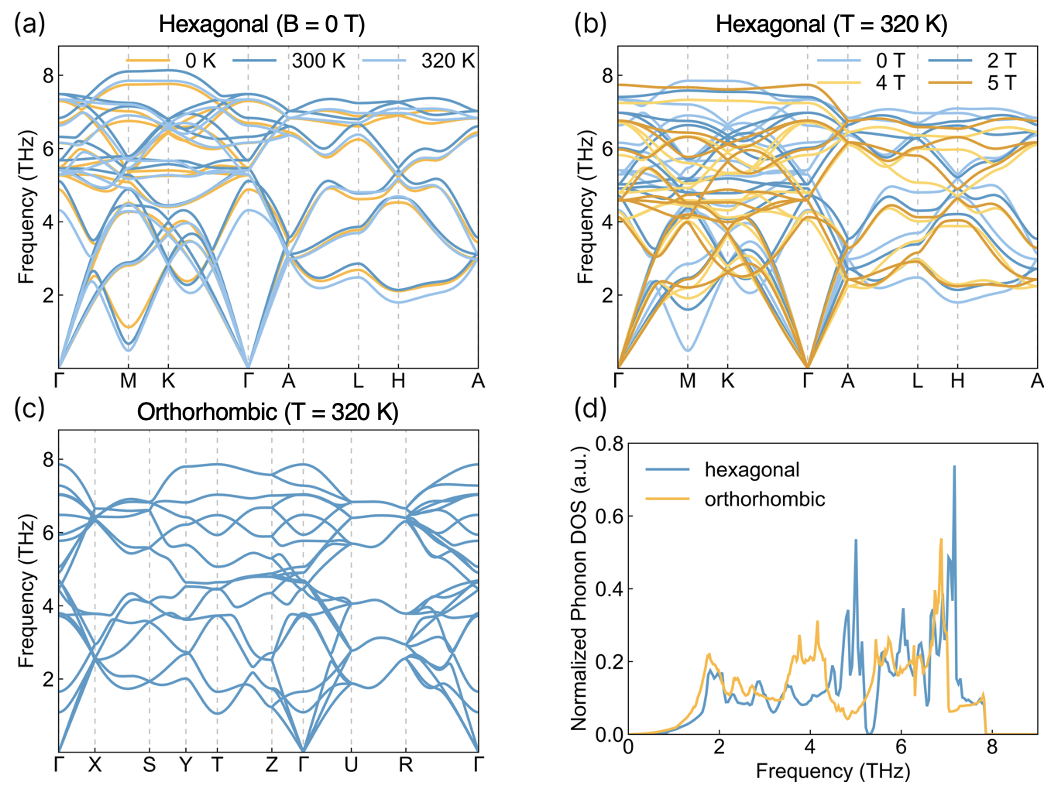}
    \caption{\textbf{Phonon properties of MnAs near the magnetostructural transition.} (a) Simulated phonon dispersions of hexagonal MnAs at zero-field as a function of temperature, showing softening of the TA phonon at $M$ as the temperature approaches $T_C$. (b) Simulated phonon dispersions of hexagonal MnAs at 320 K as a function of magnetic field, showing the hardening effect of the TA phonon by the magnetic field. (c) Simulated phonon dispersion of the orthorhombic MnAs at 320 K. (d) Simulated zero-field phonon density of states of hexagonal and orthorhombic MnAs at 320 K. }
    \label{fig:phonon}
\end{figure}

Before comparing our simulation results with experiments in detail, we further examine the origin of the lattice contribution to the total entropy change. In Fig.~\ref{fig:phonon}a, we plot the phonon dispersion of the hexagonal phase MnAs as a function of temperature from the SLD simulation, including phonon renormalization effects from both lattice anharmonicity and spin-lattice interaction. Clear softening of the transverse acoustic (TA) phonon at the $M$ point is seen as the temperature approaches $T_C$. This result is consistent with the previous theoretical picture of a soft-mode-driven phase transition~\cite{zou2008giant-ba3} and is also observed in the simulated pressure-driven phase transition~\cite{lazewski2010phonon}. Due to the first-order nature of the transition, phonon softening is incomplete, meaning that the phonon frequency does not reach zero. This soft-mode mechanism also signals that the phonon structure is highly unstable near the Curie temperature and potentially very sensitive to an applied magnetic field, which can drive MnAs away from the magnetostructural transition. This is clearly observed in Fig.~\ref{fig:phonon}b, where we show the phonon dispersion of MnAs as a function of magnetic field at 320 K, just below the simulated Curie temperature. In this case, the soft mode at $M$ is significantly hardened by the magnetic field. This large tunability of the phonon dispersion by the magnetic field is the origin of the lattice contribution to the isothermal entropy change $\Delta S_L$. The total entropy of phonons in a crystal at temperature $T$ can be calculated as~\cite{wallace2002statistical,hui1975thermodynamics}:
\begin{equation}
    S_{\mathrm{ph}}(\omega,T)=\int_0^{\omega_{\mathrm{max}}} k_{\mathrm{B}}[(n+1)\ln(n+1)-n\ln n]D_{\mathrm{ph}}(\omega)d\omega,
\end{equation}
where $\omega$ is the phonon frequency, $k_B$ is the Boltzmann constant, $n(\omega,T)$ is the equilibrium occupation number of a phonon mode, and $D_{\mathrm{ph}}(\omega)$ is the phonon density of states (DOS) at frequency $\omega$. In general, the phonon entropy increases with the occupation number $n(\omega,T)$. Therefore, phonon softening (hardening) leads to a higher (lower) occupation number and an increased (decreased) entropy. Using this formula and the phonon dispersions (thus phonon DOS) under different magnetic fields, we can compute the change in phonon entropy due to the magnetic field. In this case, the magnetic field hardens the phonons and decreases the phonon entropy, suggesting that the lattice contribution to the entropy change $\Delta S_L$ has the same sign as the magnetic contribution $\Delta S_M$. At 5 T, $\Delta S_L$ is calculated to be -9.35 J\,kg$^{-1}$\,K$^{-1}$, which agrees well with the difference between the ASD and SLD results shown above.  

There is another source of lattice contribution to the entropy change: the one associated with the structural phase transition. This will impact the isothermal entropy change $\Delta S_T$ measured at temperatures right above the Curie temperature. In this case, MnAs is initially in its orthorhombic paramagnetic state, and an applied magnetic field will drive its magnetostructural phase transition into the hexagonal phase~\cite{mira2003structural}. Intuitively, the volume expands in this process, and thus the phonons are expected to soften, leading to an increase in the phonon entropy. Therefore, the lattice entropy change associated with the structural transition, denoted $\Delta S_{PT}$ here (``PT'' stands for ``phase transition''), has an opposite sign to $\Delta S_M$ and $\Delta S_L$, canceling a portion of the total entropy change. To explicitly evaluate this contribution, we calculated the phonon dispersion of the orthorhombic phase with lattice parameters right above the Curie temperature (shown in Fig.~\ref{fig:phonon}c) and compared the phonon DOS associated with the orthorhombic phase and the hexagonal phase at zero magnetic field near $T_C$ in Fig.~\ref{fig:phonon}d. Based on the phonon DOS of the two phases, we calculated $\Delta S_{PT}$ to be $+9.64$ J\,kg$^{-1}$\,K$^{-1}$, meaning that the lattice entropy of the hexagonal phase is 9.64 J\,kg$^{-1}$\,K$^{-1}$ higher than that of the orthorhombic phase at zero magnetic field. This value matches a previous estimation based on ground state (0 K) phonon dispersions under different pressures~\cite{lazewski2010phonon}. It is intriguing to see that $\Delta S_{PT}$ has a similar magnitude (but opposite sign) as $\Delta S_L$ induced by a 5-T magnetic field. Our result suggests that the total isothermal entropy change $\Delta S_{T}$ measured right above the Curie temperature with a 5-T magnetic field is roughly -30 J\,kg$^{-1}$\,K$^{-1}$, in agreement with experimental results~\cite{wada2001giant}. Our simulation also explains the experimental observation of the sudden jump of $\Delta S_T$ by about 9 J\,kg$^{-1}$\,K$^{-1}$ when the temperature decreases below $T_C$, in which case the structural phase transition is avoided~\cite{wada2001giant}. This jump was previously attributed to an experimental artifact~\cite{wada2001giant}. Our simulation here suggests that it is likely a signature of the lattice entropy change associated with the structural phase transition.

\begin{figure}
    \centering
    \includegraphics[width=\linewidth]{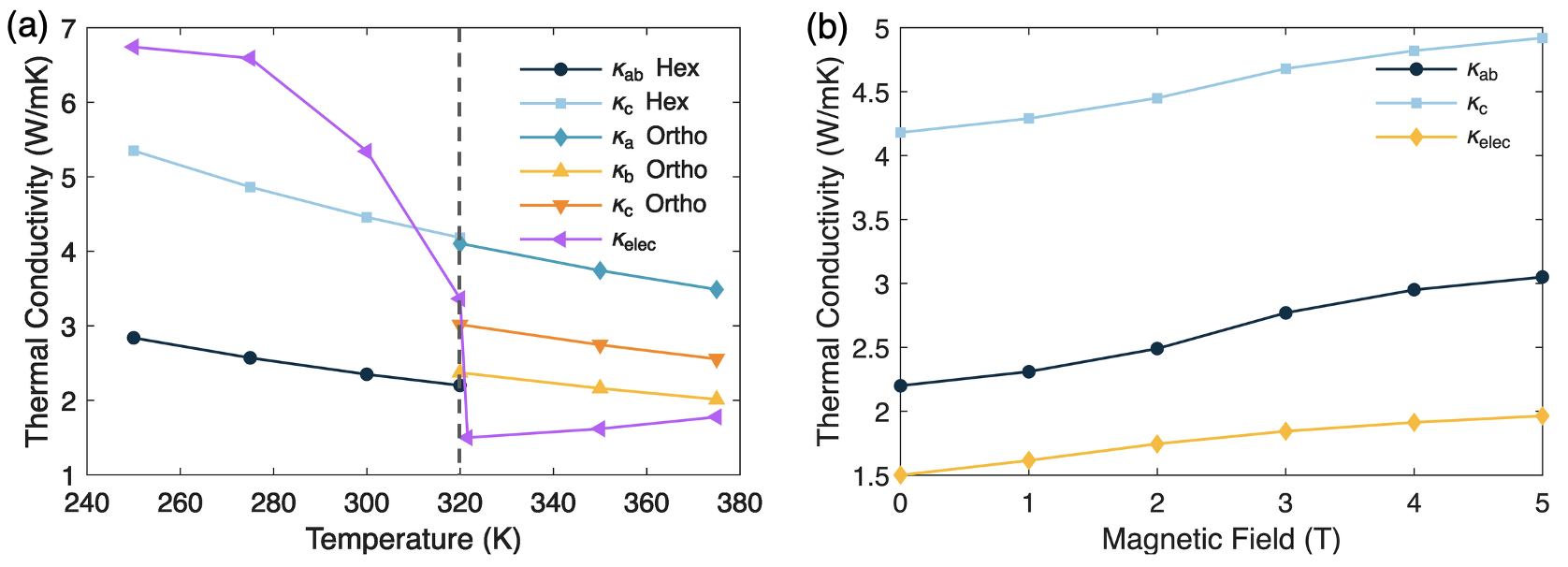}
    \caption{\textbf{Thermal conductivity of MnAs near the magnetostructural transition.} (a) Lattice thermal conductivity along different directions and the electronic thermal conductivity (``$\kappa_{\mathrm{elec}}$'') of both phases of MnAs as a function of temperature. The electronic thermal conductivity is calculated using Wiedemann-Franz law using electrical conductivity data from \cite{bean1962magnetic}. (b) Lattice thermal conductivity along different directions and the electronic thermal conductivity (``$\kappa_{\mathrm{elec}}$'') of hexagonal MnAs at 320 K as a function of magnetic field. The electronic thermal conductivity is calculated using Wiedemann-Franz law using magnetoresistance data from \cite{mira2003structural}.}
    \label{fig:thermal}
\end{figure}

Lastly, we note that the strong magnetic-field dependence of the phonon dispersion of the hexagonal phase right below $T_C$ (shown in Fig.~\ref{fig:phonon}b) indicates that its lattice thermal conductivity can be sensitively tuned by a magnetic field. In particular, hardening of the soft optical mode at $M$ by the magnetic field is expected to reduce its scattering with heat-carrying acoustic modes, thus leading to an increased lattice thermal conductivity. Given previous observation of a large negative magnetoresistance (up to -23\% with a 5-T magnetic field) in the hexagonal phase just below $T_C$~\cite{mira2003structural}, both the electronic and phononic thermal conductivity in the hexagonal phase of MnAs increase with a magnetic field, which can be useful for dynamic thermal switching applications~\cite{wehmeyer2017thermal}. With SLD and MD, the lattice thermal conductivity of the hexagonal phase and the orthorhombic phase MnAs can be extracted using the Green-Kubo formula~\cite{zhang2024room} (See more details in the Supplementary Information). Since MnAs is metallic, the electronic thermal conductivity can be estimated using the Wiedemann-Franz law based on experimental electrical resistivity data~\cite{bean1962magnetic}. Magnons can also contribute to thermal conductivity, in principle, but their contribution near room temperature is often negligible compared to that of phonons due to their much lower heat capacity~\cite{boona2014magnon}. In Fig.~\ref{fig:thermal}a, we show the calculated electronic thermal conductivity and anisotropic lattice thermal conductivity of both phases of MnAs from 250 K to 375 K across $T_C$. A slight discontinuity in the lattice thermal conductivity is observed at $T_C$, which is attributed to the contracted in-plane lattice constant. The electronic thermal conductivity decreases sharply towards the phase transition and reaches the minimum at $T_C$.
Our calculated total thermal conductivity is higher than a previous experimental report~\cite{Fujieda2004JournalofAppliedPhysics}, and the discrepancy is likely due to the fact that a sintered polycrystalline sample was measured in the experiment, where additional scattering by grain boundaries and defects can significantly reduce the thermal conductivity. In Fig.~\ref{fig:thermal}b, we show that the simulated lattice thermal conductivity along both in-plane and out-of-plane directions in hexagonal MnAs increases with the magnetic field as expected. At 320 K, this effect is maximized with approximately 39\% change along the in-plane direction and 18\% change along the out-of-plane direction. Given that hexagonal MnAs shows a large negative magnetoresistance just below $T_C$~\cite{mira2003structural}, the electronic thermal conductivity also increases with the magnetic field based on the Wiedemann-Franz law (as shown in Fig.~\ref{fig:thermal}b with electrical conductivity data from \cite{mira2003structural}). We note here that the field-dependent TA phonon frequency in hexagonal MnAs can also lead to significant change in electron-phonon interaction, which is a possible origin of the large magnetoresistence observed in MnAs~\cite{mira2003structural}. The sensitive dependence of both lattice and electronic thermal conductivity on magnetic field suggests that hexagonal MnAs near $T_C$ represents an interesting material system to explore magnetic-field-driven switching of thermal conductivity~\cite{wehmeyer2017thermal}.

\section{Conclusion}

In summary, we establish a unified microscopic picture of spin–lattice coupling across a first-order magnetostructural transition using finite-temperature first-principles spin–lattice dynamics simulations of MnAs. We show that lattice contributions to the magnetocaloric response arise from two competing mechanisms: magnetic-field–induced phonon hardening and lattice entropy changes associated with structural transformations, which have comparable magnitudes and opposite signs. This framework resolves long-standing ambiguities in lattice contributions to entropy changes in first-order magnetocaloric materials and highlights phonon tunability as a key design parameter. More broadly, our results demonstrate that engineering magnetic-field-sensitive lattice instabilities provides a general strategy for simultaneously enhancing caloric performance and enabling switchable thermal transport in multifunctional magnetic materials.

\begin{acknowledgments}
This work is partially based on research supported by the National Aeronautics and Space Administration (NASA) under award number 80NSSC21K1812 (for developing the SLD simulation) and by the Office of Naval Research under award number N00014-22-1-2262 (for potential thermal switching applications). Y.C. also acknowledges support from the NSF Quantum Foundry via the Q-AMASE-i program under award number DMR-1906325 at the University of California, Santa Barbara (UCSB). This work used Stampede2 at Texas Advanced Computing Center (TACC) through allocation MAT200011 from the Advanced Cyberinfrastructure Coordination Ecosystem: Services \& Support (ACCESS) program, which is supported by National Science Foundation grants 2138259, 2138286, 2138307, 2137603, and 2138296. Use was also made of computational facilities purchased with funds from the National Science Foundation (award number CNS-1725797) and administered by the Center for Scientific Computing (CSC) at the University of California, Santa Barbara (UCSB). The CSC is supported by the California NanoSystems Institute and the Materials Research Science and Engineering Center (MRSEC; NSF DMR-2308708) at UCSB. 
\end{acknowledgments}

\bibliography{references.bib}

\end{document}



\title{Supplementary Information: Coupled Spin-lattice Dynamics across a Magnetostructural Phase Transition} 

\author{Lokanath Patra}
\affiliation{Department of Mechanical Engineering, University of California, Santa Barbara, CA 93106, USA}

\author{Zeyu Xiang}
\affiliation{Department of Mechanical Engineering, University of California, Santa Barbara, CA 93106, USA}

\author{Yubi Chen}
\affiliation{Department of Mechanical Engineering, University of California, Santa Barbara, CA 93106, USA}
\affiliation{Department of Physics, University of California, Santa Barbara, CA 93106, USA}

\author{Bolin Liao}
\email{bliao@ucsb.edu} \affiliation{Department of Mechanical Engineering, University of California, Santa Barbara, CA 93106, USA}

\maketitle



\section{Computational Methods}
\subsection{Density Functional Theory Simulations}
\begin{figure}[!htb]
\includegraphics[width=0.7\linewidth]{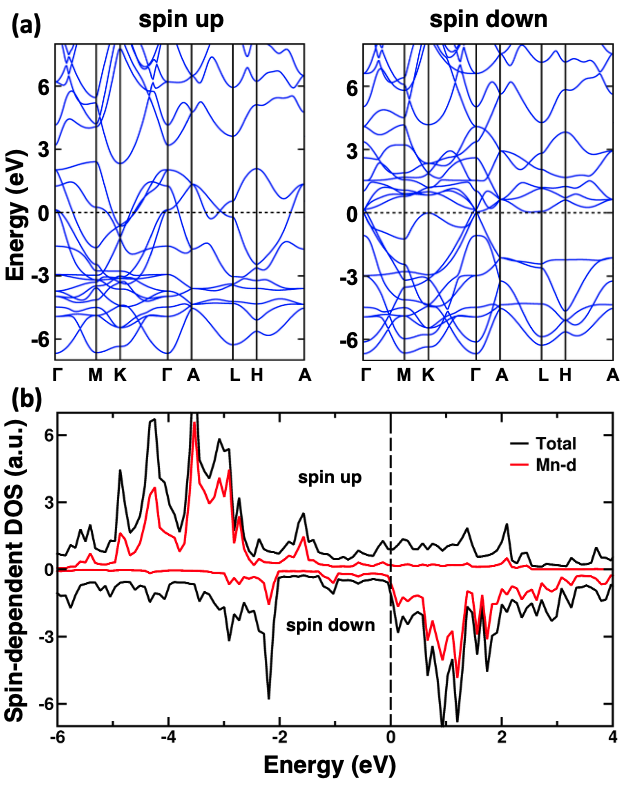}
\caption{Calculated electronic structure of hexagonal phase MnAs. (a) Spin-polarized electron band structure and (b) electronic density of states of hexagonal MnAs.} 
\label{fig:elec}
\end{figure}

Structural optimizations of the hexagonal and orthorhombic phases of MnAs were performed using density functional theory (DFT) with the Vienna Ab initio Simulation Package (VASP)~\cite{kresse1996efficient} based on the projected augmented wave (PAW) method.~\cite{blochl1994projector} The plane wave cut-off energy, energy, and force convergence criteria are set to be 600 eV, $1 \times 10^{-5}$ eV, and 0.01 eV/\AA, respectively. The Perdew–Burke–Ernzerhof form of the generalized gradient approximation (PBE-GGA)~\cite{perdew1996generalized} was employed for the exchange-correlation functional in all calculations. The Monkhorst-Pack~\cite{monkhorst1976special} \textbf{k}-point meshes of $12 \times 12 \times 8$ and $12 \times 8 \times 8$ were used for hexagonal and orthorhombic phases, respectively. The calculated spin-polarized electronic structure is shown in Fig.~\ref{fig:elec}.

\begin{figure}[!htb]
\includegraphics[width=0.6\linewidth]{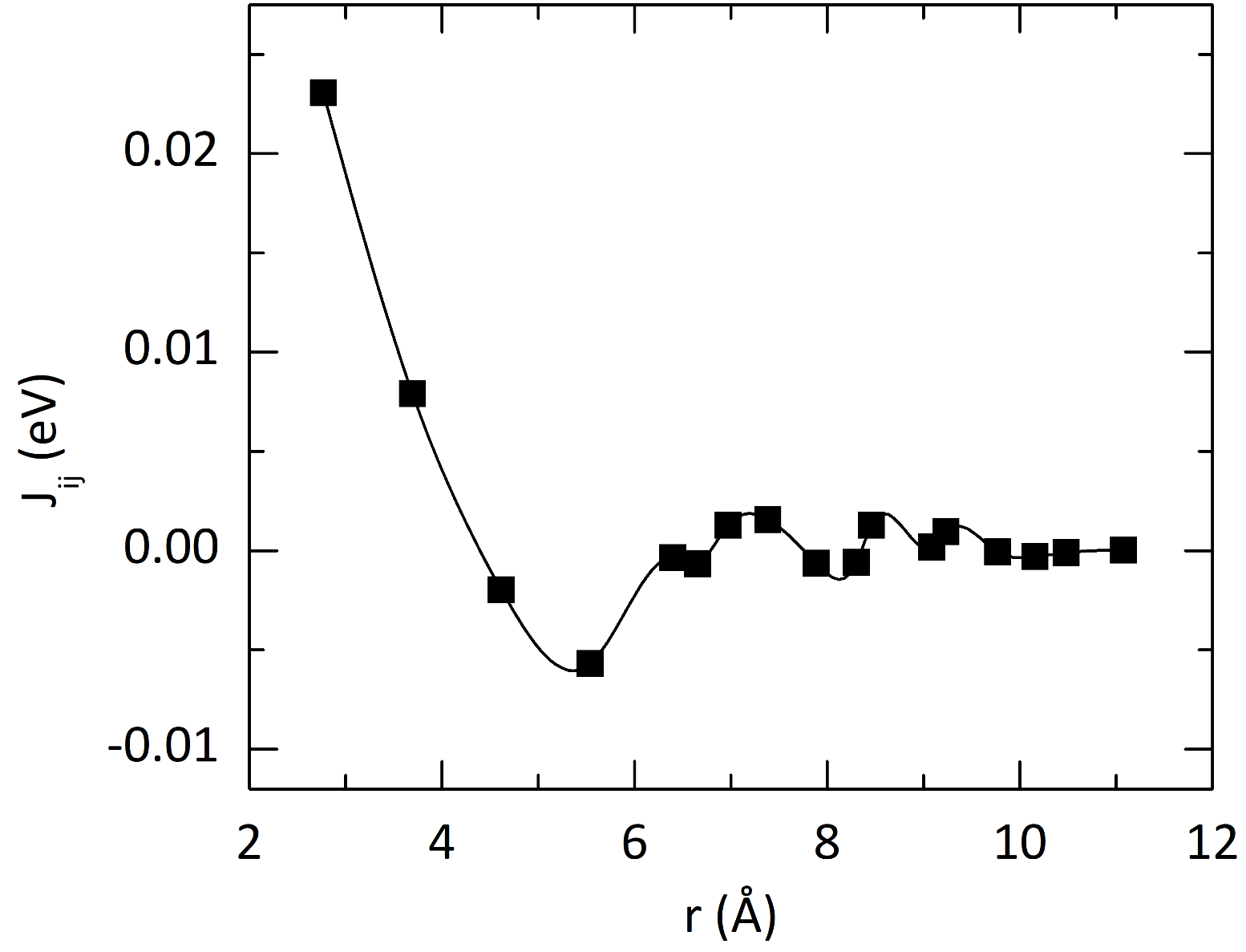}
\caption{Calculated spin exchange parameters in hexagonal MnAs as a function of atomic distance.} 
\label{fig:spin}
\end{figure}

The full-potential linear muffin-tin orbital method (FP-LMTO) was employed within the SPR-KKR package~\cite{ebert2011calculating} to calculate the magnetic exchange parameters, $J_{ij}$. These parameters were extracted by fitting a Heisenberg Hamiltonian to the ground-state energies calculated for different spin configurations: 
\begin{equation}
    H_h = -\sum_{i \neq j} J_{ij} \mathbf{S}_i \cdot \mathbf{S}_j,
\end{equation}
where $\mathbf{S}_i$ and $\mathbf{S}_j$ are unit vectors pointing in the direction of local magnetic moments at atomic sites $i$ and $j$. The calculated $J_{ij}$ in hexagonal MnAs as a function of interatomic distance is shown in Fig.~\ref{fig:spin}.

\subsection{First Principle Calculation of U}
Hubbard $U$ of Mn $d$-electrons is calculated using the linear response approach~\cite{cococcioni2005linear}, in which $U$ is derived from the self-consistent response function ($\chi$) and the non-self-consistent response function ($\chi_0$):
\begin{equation}
   U=\chi^{-1}-\chi_0^{-1}\approx \left(\frac{\partial N_I^{\mathrm{SCF}}}{\partial V_I}\right)^{-1}-\left(\frac{\partial N_I^{\mathrm{NSCF}}}{\partial V_I}\right)^{-1}, 
\end{equation}
where $V_I$ is the localized potential shift applied to atom $I$, $N_I^{\mathrm{SCF}}$ and 
$N_I^{\mathrm{NSCF}}$ are the resulting self-consistent and non-self-consistent electron occupation number, respectively. To ensure the validity of the linear-response assumption, a $3\times3\times2$ supercell is employed to apply localized perturbations of potential shifts on one Mn $d$ orbital. 
The applied potential shifts are constrained within the range of $-0.2$~eV to $0.2$~eV. The estimated Hubbard U value of 5.15 eV was adopted for Mn in MnAs in all subsequent calculations.

\begin{figure}[!tb]
\includegraphics[width=0.5\linewidth]{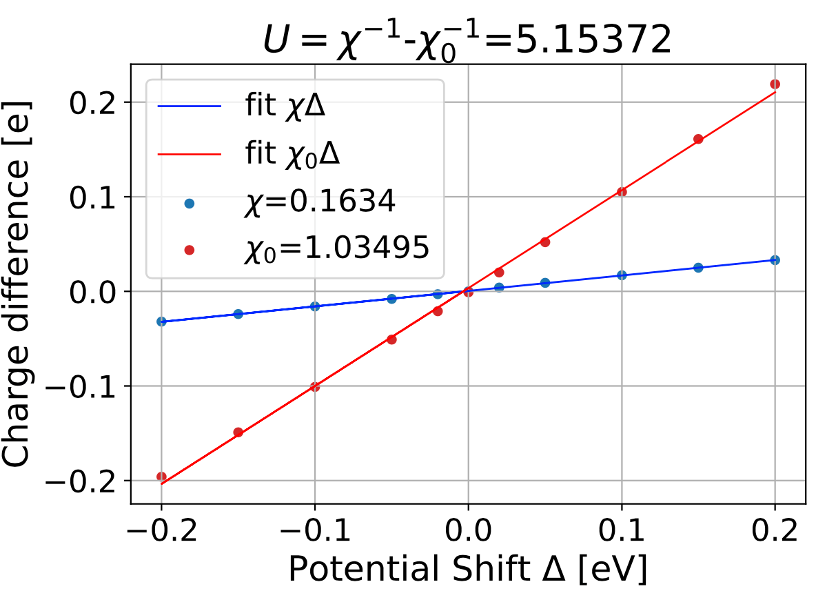}
\caption{First-principles determination of the Hubbard $U$ parameter for Mn in MnAs. } 
\label{fig:uval}
\end{figure}

\subsection{Spin Lattice Dynamics Simulations}
The calculated magnetic exchange parameters serve as inputs for the spin lattice dynamics (SLD) simulations performed in this work. These simulations were carried out using the SPIN package within the LAMMPS software~\cite{tranchida2018massively}, which introduces magnetic effects into classical molecular dynamics through a generalized Hamiltonian, i.e.: 

\begin{equation}
    \mathcal{H} = \sum_{i=1}^{N}\frac{|\textbf{\textit{p}}_i|^2}{2m_i} + \sum_{i, j, i\neq 1}^{N} V (r_{ij}) + \mathcal{H}_{mag},
\end{equation}
where the first two terms represent the kinetic energy and the classical interatomic potential, respectively, that together describe the mechanical motion of the atoms. The last term, $\mathcal{H}_{mag}$, is the magnetic Hamiltonian. Although it can account for complex effects such as magnetic anisotropy, dipolar, Dzyaloshinskii-Moriya, and magnetoelectric interactions, this study focuses solely on spin-spin exchange interactions and interactions with external magnetic fields ($\mathcal{H}_{ext}$). The simplified magnetic Hamiltonian used in the present work is given by:

\begin{equation}
    \mathcal{H}_{mag} = \sum_{i, j, i\neq 1}^{N} \textbf{\textit{J}}(r_{ij})\textbf{\textit{S}}_i \cdot \textbf{\textit{S}}_j + \mathcal{H}_{ext},
\end{equation}
where $J(r_{ij})$ represents the magnetic coupling exchange constant as a function of the interatomic distance ($r_{ij}$) between atoms $i$ and $j$, and \textbf{\textit{S}}$_i$ and \textbf{\textit{S}}$_j$ are the normalized spin vectors of spin $i$ ($j$). For this study, only the magnetic exchange interactions between Mn atoms were taken into account; interactions involving As (i.e., Mn-As and As-As pairs) were neglected. The calculated exchange parameters were fitted to the following Bethe–Slater curve:
\begin{equation}
    J(R_{ij}) = 4a\left(\frac{R_{ij}}{d}\right)^2\left(1-b\left(\frac{R_{ij}}{d}\right)^2\right)e^{-\left(\frac{R_{ij}}{d}\right)^2}\Theta\left(R_c-R_{ij}\right),
\end{equation}
where $a$, $b$, and $d$ are fitting parameters and $R_c$ is the cutoff radius associated with the pair interaction.

SLD calculations were performed using a supercell size of $20 \times 20 \times 20$ with the recently included SPIN package of the LAMMPS simulation tool. The Mn-As EAM potential developed by Andreas R{\"u}hl was used in our simulations~\cite{Ruehl2016}. The initial direction of the spins was set along the $c$-direction. Next, the systems were relaxed for 50 ps with a time step of 1 fs at 300 K and 0 bar. A Langevin thermostat and a Berendsen barostat were utilized to control the temperature and maintain the crystal at zero pressure, respectively. Temperature dependence analyses were performed by heating the crystal to the desired temperature and then equilibrating it for 10 ps under the application of the zero-pressure barostat. 

The field-dependent $\Delta S_T$ values were estimated from the magnetization vs. temperature relations using Maxwell’s relation (Eqn. 1 in the main text).
The FixPhonon command within the LAMMPS package was utilized to calculate the phonon spectra~\cite{kong2011phonon}, which is based on measuring the correlations of atomic displacements in the SLD and MD simulations. The simulations were conducted in the constant volume-temperature (NVT) ensemble at the desired temperatures, with the volume for each temperature set to its optimized value. A supercell size of $20 \times 20 \times 20$ was utilized for this purpose. The obtained dynamic matrices calculated by FixPhonon were subsequently analyzed with the ``phana'' post-processing code to obtain phonon dispersion and phonon density of states. To obtain the lattice thermal conductivity, the Green-Kubo molecular dynamics simulations were performed using the LAMMPS package. A 50 ps NVT simulation, followed by a 50 ps NVE simulation, was performed to stabilize the system. The calculation of thermal conductivity was then performed for a period of 1 ns. 

The total thermal conductivity was calculated as the sum of the electronic and phonon contributions:
\begin{equation}
    k_{tot} = k_e + k_{ph},
\end{equation}
where $k_e$ is the electronic thermal conductivity and $k_{ph}$ is the phonon thermal conductivity. The electronic component, $k_e$, is calculated using the Wiedemann-Franz law:
\begin{equation}
    k_e = \frac{LT}{\rho}
\end{equation}
where $L$, $T$, and $\rho$ represent the Lorenz number, absolute temperature, and electrical resistivity, respectively. The electrical resistivity data were collected from experimental measurements reported by Bean \textit{et al}.~\cite{bean1962magnetic}.

\clearpage

\section{Additional Data}

\clearpage

\clearpage

\begin{figure}[!htb]
\includegraphics[width=0.7\linewidth]{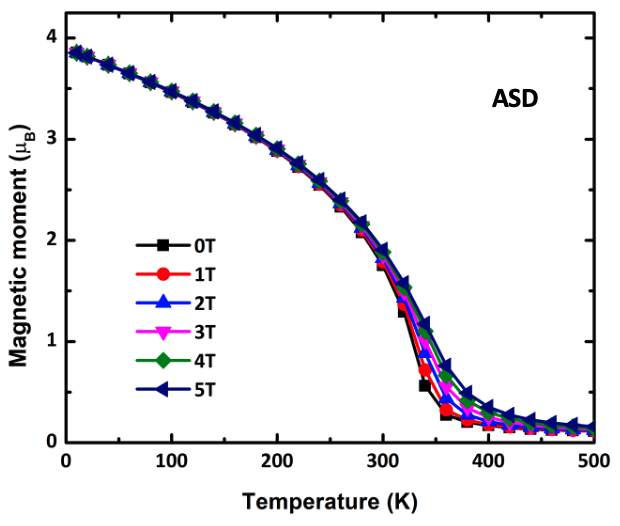}
\caption{Calculated magnetization as a function of temperature and magnetic field in hexagonal MnAs using ASD.} 
\label{fig:ASD}
\end{figure}

\clearpage

\begin{figure}[!htb]
\includegraphics[width=0.7\linewidth]{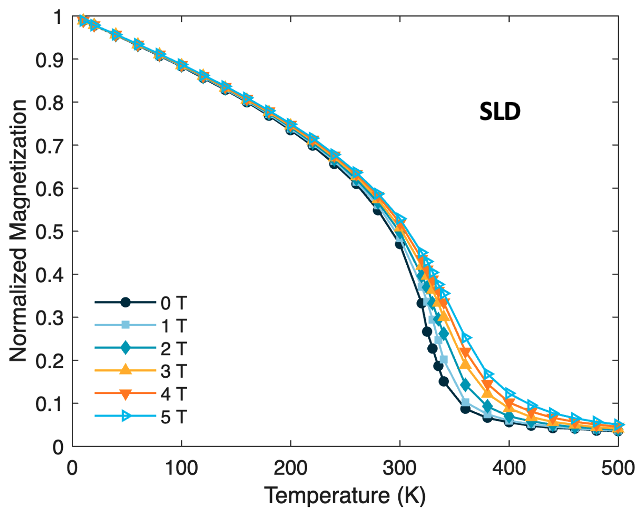}
\caption{Calculated magnetization as a function of temperature and magnetic field in hexagonal MnAs using SLD.} 
\label{fig:SLD}
\end{figure}
\clearpage

\section*{Supplementary References}
\bibliography{references.bib}